\documentclass[12pt,a4paper]{article}
\usepackage{eurosym}
\usepackage{amssymb}
\usepackage[toc,page]{appendix}

\input{tcilatex}
\begin{document}

\date{}
\title{An analogy between optical turbulence and activator-inhibitor dynamics%
}
\author{F. Spineanu and M. Vlad \\
National Institute of Laser, Plasma and Radiation Physics\\
Magurele, Bucharest 077125, Romania}
\maketitle

\begin{abstract}
The propagation of laser beams through madia with cubic nonlinear
polarization is part of a wide range of practical applications. The
processes that are involved are at the limit of extreme (cuasi-singular)
concentration of intensity and the transversal modulational instability, the
saturation and defocusing effect of the plasma generated through avalanche
and multi-photon (MPI) ionization are competing leading to a complicated
pattern of intensity in the transversal plane. This regime has been named
\textquotedblleft optical turbulence\textquotedblright and it has been
studied in experiments and numerical simulations. Led by the similarity of
the portraits we have investigated the possibility that the mechanism that
underlies the creation of the complex pattern of the intensity field is the
manifestation of the dynamics \textit{activator-inhibitor}. In a previous
work we have considered a unique connection, the \textit{complex
Landau-Ginzburg equation}, a common ground for the nonlinear Schrodinger
equation (optical propagation) and reaction-diffusion systems
(activator-inhibitor). The present work is a continuation of this
investigation. We start from the exact integrability of the elementary
self-focusing propagation (\textit{gas Chaplygin with anomalous polytropic})
and show that the analytical model for the intensity can be extended on
physical basis to include the potential barrier separating two states of
equilibria and the drive due to competing Kerr and MPI nonlinearities. We
underline the variational structure and calculate the width of a branch of
the cluster of high intensity (when it is saturated at a finite value). Our
result is smaller but satisfactorily in the range of the experimental
observations.
\end{abstract}

\tableofcontents

\pagestyle{empty}

\section{Introduction}

This work is an extension of our previous work on the possible parallel
between the optical turbulence and the \emph{Labyrinth instability} acting
in a system with a dynamics of the type \emph{activator-inhibitor} \cite%
{florinmadilabyrinth}. We recall that \emph{optical turbulence} is one of
the regimes of propagation in a medium with cubic Kerr nonlinearity of a
pulse produced by a laser at powers much higher than the threshold for
self-focalization. The multiple filamentation, saturation through generation
of plasma followed by re-location and coalescence of zones of high intensity
lead to a complicated distribution of intensity in the transversal plane.
The basic mechanism for the apparently random distribution is similar to a
competition of two fields in a reaction-diffusion system. One is
auto-catalitic and the other acts to limit the expansion of the first.
Previously we have argued that a connection can be established between the
analytical structure underlying the optical turbulence and the one of the
labyrinth instability. The connection is provided by the \emph{complex
Landau-Ginzburg equation} for which exist mappings to the Nonlinear
Schrodinger Equation and respectively to activator-inhibitor equations.

In the present work we start from the description of the self-focusing as an
exactly integrable \textquotedblleft Chaplygin gas with anomalous polytropic
exponent\textquotedblright\ (or: \textquotedblleft
drop-on-ceil\textquotedblright\ \cite{trubnikovzhdanov}). We extend this
pure self-focusing scheme by adding analytical terms which are manifestation
of natural physical processes:

\begin{itemize}
\item the diffusion

\item the difference in potential energy between the two extrema at
equilibrium: $I=I_{\max }$ and $I=0$;

\item the competition between Kerr nonlinearity and the defocusing property
of the plasma
\end{itemize}

Therefore we must note from the beginning that the theory is constructed on
the basis of analytical implementation of properties that are identified in
a physical analysis of the two real systems.

We show (Appendix  \ref{App:AppendixA}) that a modification of the exactly
integrable \textquotedblleft drop-on-ceil\textquotedblright\ instability
exhibits the expected effect of increasing structuring in the transversal
plane

We study the possible stabilization of the width of a stripe belonging to
the cluster of high intensity. For the range of parameters that permit
stabilization, we can provide an approximative value. \ Compared with
experimental observation, our analytical result is smaller, but the sources
of improvement of the analytical approach are sufficiently rich to allow
extensions.

\subsection{The basic analytical model of the propagation with self-focusing}

We start from the basic elements of the propagation of a high intensity
laser pulse in a cubic nonlinear medium. Consider the equation for the
amplitude of the electric field $A\left( z,x,y\right) $ of a laser beam $%
\left( k_{0},\omega _{0}\right) $ in a medium with Kerr nonlinearity $%
\varepsilon _{2}>0$, 
\begin{equation}
2ik_{0}\frac{\partial A}{\partial z}+\Delta _{\perp }A+k_{0}^{2}\frac{%
\varepsilon _{2}}{\varepsilon _{0}}\left\vert A\right\vert ^{2}A=0
\label{eq1}
\end{equation}%
and take a new factorization, in which it is introduced the \emph{eikonal}%
\begin{equation}
A\left( r,z\right) =a\left( r,z\right) \exp \left[ ik_{0}S\left( r,z\right) %
\right]   \label{eq2}
\end{equation}%
where $S\left( r,z\right) \equiv $ eikonal with unit $\left[ S\right] =$%
length. The resulting equations are (\cite{Talanov1965}, \cite{schvartsburg1}%
, \cite{trubnikovzhdanov}), assuming axial symmetry in the transversal plane
(\emph{i.e.} only retaining the radial coordinate $r$) 
\begin{eqnarray}
\frac{\partial a}{\partial z}+v\frac{\partial a}{\partial r}+\frac{a}{2r}%
\frac{\partial }{\partial r}\left( rv\right)  &=&0  \label{eq3} \\
2\frac{\partial S}{\partial z}+v^{2} &=&\frac{\varepsilon _{2}}{\varepsilon
_{0}}a^{2}+\frac{1}{k_{0}}\frac{\Delta _{\perp }a}{a}  \nonumber
\end{eqnarray}%
where the \textquotedblleft velocity\textquotedblright is%
\begin{equation}
v=\frac{\partial S}{\partial r}  \label{eq4}
\end{equation}%
nondimensional. The \emph{velocity} is the derivative of the eikonal to the
radial coordinate. It actually is like a \emph{wavenumber} for a propagation
in the \emph{transversal }direction to the $z$ axis. It will govern the
pattern formation in the \emph{transversal} plane. The last term can be
neglected in the limit $\lambda \rightarrow 0$. Then, adopting the new
variable%
\begin{equation}
I\equiv a^{2}  \label{eq5}
\end{equation}%
we have%
\begin{eqnarray}
\frac{\partial I}{\partial z}+\frac{1}{r}\frac{\partial }{\partial r}\left(
rvI\right)  &=&0  \label{eq6} \\
\frac{\partial v}{\partial z}+v\frac{\partial v}{\partial r} &=&\frac{%
\varepsilon _{2}}{2\varepsilon _{0}}\frac{\partial I}{\partial r}  \nonumber
\end{eqnarray}%
These equations are of type \textquotedblleft drop-on-ceil instability\textquotedblright and belong to the
class describing \emph{a gas Chaplygin with anomalous politropic exponent}.
They can only be solved approximately. To advance the analytical description
it is necessary to restrict to a single spatial coordinate in the
transversal plane, which renders the system exactly integrable%
\begin{eqnarray}
\frac{\partial I}{\partial z}+\frac{\partial }{\partial x}\left( vI\right) 
&=&0  \label{eq7} \\
\frac{\partial v}{\partial z}+v\frac{\partial v}{\partial x} &=&c_{0}\frac{%
\partial }{\partial x}\left( \frac{I}{I_{0}}\right)   \nonumber
\end{eqnarray}%
Here%
\begin{equation}
c_{0}^{2}=\frac{\varepsilon _{2}}{2\varepsilon _{0}}I_{0}  \label{eq8}
\end{equation}%
and $I_{0}=a_{0}^{2}$ is the intensity at the entrance in the medium. These
equations are solved in Appendix  \ref{App:AppendixA} using the hodograph
transformation, as described in \cite{trubnikovzhdanov}.

\bigskip

\subsection{The optical turbulence}

To investigate the possible validity of the parellel between optical
turbulence and the activator-inhibitor dynamics we will not employ a
detailed description of the random multiple filamentation pattern of
intensity. We must retain that there are regions of high intensity and
complementary regions of low intensity. Their spatial pattern is an
intricate distribution of stripes (branches of a plane graph) as connected
components of a cluster. Further we will mention that inside the regions of
the cluster of high intensity there are spots of even higher intensity,
where new filaments are initiated. This is because the intensity is still
higher than the threshold for self-focusing. In such a spot it is generated
plasma and the effect of the electrons of the plasma is to defocus locally
the beam and to saturate the increase of the\ intensity. This is seen as a
relocation of the high intensity from the region of concentration. We then
recognize the basic dynamics of an activator with auto-catalitic evolution
(the intensity) and a competing inhibitor (the plasma).

The sequence of physical processes is as follows: (1) The high intensity produced at self-focalization generates plasma; (2) Plasma acts as a negative lens.; (3) Plasma pushes away the high intensity spots while it expands and
de-localizes them. (This has experimental support: in a symmetric geometry 
\cite{tzortzakis} the axial region of the high-intensity pulse is moved
symmetrically towards larger radii and a ring is formed. No substantial loss
of energy occurs at these events. Then the ring collapses again on the axis.)

This is the physical picture that we will have to implement in an analytical
description.

\bigskip

\section{Expanding around the strict self-focusing dynamics}

We will draw a parallel between the optical turbulence and the dynamics of
an \emph{activator-inhibitor} system. With only the Kerr nonlinearity
retained, the equation for the intensity%
\begin{equation}
\frac{\partial I}{\partial z}+\frac{\partial }{\partial x}\left( vI\right) =0
\label{eq9}
\end{equation}%
is an equation of conservation where the effect of advection is produced by the transversal variation of the eikonal. The focusing effect creates in the
transversal plane regions where the intensity $I$ is high while in the
complementary zone $I$ is relatively low (see Ettoumi \emph{et al.} \cite%
{phasetransition}). As suggested by the approach in the case of
reaction-diffusion systems, we will simplify the representation of the
intensity field by restricting it to only two values : $I=I_{\max }$ and
respectively $I=0$, uniformly distributed inside mutually excluded zones 
\cite{goldsteinlabyrinth}. These zones are stripes with meandering shapes in
plane, each creating a connected cluster and separated by sharp interfaces
(as in Fig.1 of Ref.\cite{phasetransition}) from the complementary set. The
evolution of the system from one state to another is constrained. This means
that in a point $x$, through only successive steps consisting of focusing,
plasma generation by ionization, defocusing and relocation of high-$I$
regions there can be transition from one state to another. This
particularity is very often encountered (including to reaction-diffusion
systems) and is represented schematically as a potential with two
equilibrium states separated by a barrier%
\begin{equation}
F\left[ I\right] \sim I^{2}\left( I-I_{\max }\right) ^{2}  \label{eq10}
\end{equation}%
To solve Eq.(\ref{eq9}) we must find $v\left( z,x\right) $, \emph{i.e.} find
from Eq.(\ref{eq6}) the characteristics of the cuasi-Lagrangian flow of $I$.
However we would like to include at least a schematic description of the
complex processes mentioned above: focalization, plasma generation and
defocusing with re-location. Then we return to the Eulerian point of view by
assuming that changes of $I$ from $I=0$ to $I=I_{\max }$ result from the
competition between the potential energy $F$ and the external nonlinear
drive, \emph{i.e.} the Kerr focalization and the coupling with the plasma
density. The flux%
\begin{equation}
\Gamma _{I}=vI=-D\frac{\partial I}{\partial x}  \label{eq11}
\end{equation}%
ensure that the profiles are smooth. The external nonlinear drive arises
from the difference between the Kerr-induced focalization and the defocusing
effect of the density $\rho $ of electrons of the plasma, at the current
value of the intensity $I$.

The structure of alternating stripes of high intensity and zones of low
intensity (from where the high intensity has been pushed away and relocated)
appears in experiments and in numerical simulations of multiple
filamentation and optical turbulence \cite{mlejnekturbulence}, \cite%
{mlejnek98}. We are interested in the dynamics of a $x-$interval, a section
of a stripe of high intensity $I$ bounded (to the left and right) by zones
of low intensity. The high $I$ is necessarily associated with presence of
electron plasma $\rho $. In activator-inhibitor dynamics the fronts of the
activator $\left( I\right) $ are sharp while the profiles of $\rho $
(inhibitor) are expected to be smooth and diffuse. We want to see if a
stripe of high-$I$ is stabilized to a finite width limited by the left and
right fronts.

In the regions of high intensity new spots of focalization are initiated
with the tendency of formation of high concentration and further
filamentation. They are visible for example in Fig.5 of Ref.\cite{mechain3}.
Since such a spot produces plasma with defocusing and re-location effect,
one concludes that these are the positions where the modification of the
interface takes place. The two factors: activator $\left( I\right) $ and
inhibitor $\left( \rho \right) $ are always connected and $\rho $ follows $I$%
. The result is that behind their permanent competition there remain zones
with low values of both $I$ and $\rho $.

As discussed above, this complex process manifests itself as a barrier that
makes the two equilibria states to be separated and not easily mutually
accessible. It is represented by the potential $F$ with the two equilibria
states and the barrier between them. We now must postulate that the two
states of equilibrium have different {\it potential energy}, one of the states
being favored: the mix of high intensity trying to focus but saturated
through the effect of $\rho $ has higher potential energy than the empty
regions which only remain behind such events. The difference is measured as 
\cite{goldsteinlabyrinth} 
\begin{equation}
F\left[ I\right] =f\frac{1}{4}\overline{i}^{2}\left( \overline{i}%
^{2}-1\right) +\left( r-\frac{1}{2}\right) \left( \frac{1}{2}\overline{i}%
^{2}-\frac{1}{3}\overline{i}^{3}-\frac{1}{12}\right)  \label{eq12}
\end{equation}%
with $\overline{i}\equiv \frac{I}{I_{\max }}$ and $f$ is a dimensional
factor. The drive produced on the variable $I$ is%
\begin{equation}
\frac{\delta F\left[ I\right] }{\delta I}=f\frac{1}{I_{\max }^{3}}I\left(
I-rI_{\max }\right) \left( I-I_{\max }\right)  \label{eq13}
\end{equation}%
The difference between the potential energy of the two equilibrium states is 
\begin{eqnarray}
\Delta F &=&F\left[ \overline{i}=1\right] -F\left[ \overline{i}=0\right]
\label{q14} \\
&=&f\frac{1}{6I_{\max }^{3}}\left( r-\frac{1}{2}\right)  \nonumber
\end{eqnarray}%
for $0<r<1$. Now regarding the source of local dynamics, we note that the
change from one state to another can be done when there is no compensation
between Kerr focusing and plasma defocusing. The terms arise from the
substraction: $\sim $ (Kerr focusing)\ $-$\ (plasma defocusing), as in the
original extended NSEq \cite{Couaironfilamentation}, \cite{bergephysrep} 
\begin{equation}
2ik_{0}\frac{\partial E}{\partial z}\sim \frac{2k_{0}\omega _{0}}{c}%
n_{2}\left\vert E\right\vert ^{2}E-k_{0}\omega _{0}\sigma \tau _{0}\ \rho \ E
\label{eq15}
\end{equation}%
Then the coupling $C$ that acts like a drive is the difference, after
factoring out $k_{0}$, can be written%
\begin{equation}
C\equiv \alpha I^{2}-\alpha ^{\prime }I\rho  \label{eq16}
\end{equation}%
where%
\begin{equation}
\alpha \equiv \frac{2\omega _{0}}{c}n_{2}\ \ \textrm{and}\ \ \alpha ^{\prime
}\equiv \omega _{0}\sigma \tau _{0}  \label{eq17}
\end{equation}%
This coupling is no more linear as it was in classical activator-inhibitor
models \cite{desaikapral}, like FitzHugh-Nagumo.

\section{The dynamics of the stripes of intensity}

The basic analytical structure of the self-focusing instability is captured
by the {\it drop-on-ceil} instability, Eq.(\ref{eq7}). As discussed before this
structure is now extended by adding the terms representing the potential
energy cost of moving between the two distinct equilibria and by the drive
resulting from the competition of the focusing and defocusing effects. We
propose the equation 
\begin{equation}
\frac{\partial I}{\partial z}=D\frac{\partial ^{2}I}{\partial x^{2}}-\frac{%
\delta F}{\delta I}+\alpha I^{2}-\alpha ^{\prime }\rho I  \label{eq18}
\end{equation}%
after replacing the flux $\Gamma =-D\frac{\partial I}{\partial x}$. The
coordinate $x$ is measured across the section of connex stripes. The Eq.(\ref%
{eq18}) can be derived from the functional%
\begin{equation}
\mathcal{W}_{I}=\int dx\left[ \frac{1}{2}\left( \frac{\partial I}{\partial x}%
\right) ^{2}+F\left[ I\right] -\alpha \frac{I^{3}}{3}\right] +\alpha
^{\prime }\frac{1}{2}\int dx\rho \left( x\right) I^{2}\left( x\right)
\label{eq19}
\end{equation}

\bigskip

\section{The equation for the electron plasma density}

The equation for $\rho $ is \cite{Couaironfilamentation}, \cite{bergephysrep}%
, \cite{mlejnekturbulence}%
\begin{equation}
\frac{\partial \rho }{\partial t}=d\frac{\partial ^{2}\rho }{\partial x^{2}}%
-a\rho ^{2}+bI^{K}  \label{eq20}
\end{equation}

The first term in the RHS is the divergence of the local flux of density, 
\emph{i.e.} the accumulation or depletion of density, the second is the
decrease of the density through recombination and the source of density is
the last term (\textbf{note} that we have neglected the avalanche ionization 
$\sim \rho I$, which may be justified in the case of short time of pulse).
The last term is the Multi-Photon Ionization (MPI) rate.

We will investigate the state where stripes of constant $I=I_{\max }$
alternate with stripes of low intensity, $I=0$. Then we consider that the
intensity has no spatial variation and the equation of $\rho $ can be solved
with constant and uniform $I^{K}$.%
\begin{equation}
I^{K}=\textrm{const}  \label{eq21}
\end{equation}%
The parameter $d\equiv \delta ^{2}/\tau \equiv $ diffusion coefficient of
electrons\ $\left( \frac{m^{2}}{s}\right) $ is estimated in the Appendix 
\ref{App:AppendixB}. We choose%
\begin{equation}
\delta \sim 10^{-6}\ \ \left( m\right)   \label{eq22}
\end{equation}%
which is a reasonable choice in the range of possible lengths of the
electron mean free path. Using $\tau \sim 1\times 10^{-13}\ \left( s\right) $%
\ \cite{mlejnek98} we obtain%
\begin{equation}
d\sim \frac{10^{-12}}{10^{-13}}=10\ \ \left( \frac{m^{2}}{s}\right) 
\label{eq24}
\end{equation}%
Other paramaters are $a=5\times 10^{-13}\ \left( \frac{m^{3}}{s}\right) $ and%
\begin{equation}
\beta ^{\left( K=7\right) }=6.5\times 10^{-104}\ \left( \frac{m^{11}}{W^{6}}%
\right)   \label{eq25}
\end{equation}%
leading to%
\begin{equation}
b\equiv \frac{\beta ^{\left( 7\right) }}{K\hslash \omega _{0}}=3.6\times
10^{-86}\ \ \ \left( \frac{m^{11}}{J}\right)   \label{eq26}
\end{equation}%
and%
\begin{equation}
E^{phys}=9.15\times 10^{7}\ \left( \frac{V}{m}\right)   \label{eq28}
\end{equation}%
In terms of intensity we have $I\equiv \left\vert \widetilde{E}%
_{0}\right\vert ^{2}\ \ \left( \frac{W}{m^{2}}\right) $ alternatively\ \ \ \ 
$I=\left( \left\vert \sqrt{c\varepsilon _{0}}E^{phys}\right\vert ^{2}\right) 
$ where$\ \ \widetilde{E}=5\times 10^{6}\ \left( \frac{W^{1/2}}{m}\right) $
such that, calculated below, we have for $K=7$%
\begin{equation}
bI^{K}=\frac{\beta ^{\left( K=7\right) }}{K\hslash \omega _{0}}\left\vert 
\widetilde{E}_{0}\right\vert ^{2}=\frac{\beta ^{\left( K=7\right) }}{%
K\hslash \omega _{0}}\left\vert \sqrt{c\varepsilon _{0}}E_{0}\right\vert
^{2K}\sim 7.7\times 10^{7}\ \left( \frac{1}{m^{3}s}\right)   \label{eq29}
\end{equation}

It is interesting to estimate the density $\rho $ that results if the only
process were recombination $\partial \rho /\partial t=\left\vert a\rho
^{2}\right\vert $. Taking the time duration of the pulse $\delta t=80\
\left( fs\right) $ we have the estimation $1/\rho =a\times \delta t$ or $%
\rho \sim 2\times 10^{25}\ \left( part/m^{3}\right) $. On the other hand one
expects that the plasma density is approximately $1\%$ of the air
density. Then for various estimations we take%
\begin{equation}
\rho \sim 10^{23}\ \left( m^{-3}\right)  \label{eq30}
\end{equation}

\bigskip

Further, the equation can be integrated once%
\begin{equation}
d\frac{1}{2}\left( \frac{\partial \rho }{\partial x}\right) ^{2}=a\frac{1}{3}%
\rho ^{3}-b\rho I^{K}+C  \label{eq34}
\end{equation}%
where%
\begin{equation}
\left[ C\right] =\frac{1}{m^{6}s}  \label{eq35}
\end{equation}%
If the spot is symmetric the density created by $I^{K}$ has a maximum at the
center of the spot and%
\begin{eqnarray}
\frac{\partial \rho }{\partial x} &=&0\ \ \textrm{for }x=0  \label{eq36} \\
C &=&\rho \left( 0\right) bI^{K}-\frac{1}{3}a\left[ \rho \left( 0\right) %
\right] ^{3}  \nonumber
\end{eqnarray}%
We will use the notation $\rho _{0}\equiv \rho \left( 0\right) $. Replacing
in the right hand side%
\begin{equation}
\frac{d\rho }{\left[ \frac{2a}{3d}\left( \rho ^{3}-\rho _{0}^{3}\right) -%
\frac{2b}{d}I^{K}\left( \rho -\rho _{0}\right) \right] ^{1/2}}=\pm dx
\label{eq37}
\end{equation}%
We recall that we look for a regime of fast \emph{inhibitor} \cite%
{goldsteinlabyrinth}\textbf{. }This setting of the problem assumes that
there is \emph{no time variation} of the density, in the sense that the
formation of plasma is instantaneous under the effect of $I^{K}$. Only
spatial variation of the electron density is considered. Then from a
reference value of $\rho $, denoted $\rho \left( 0\right) $ at $x=0$ all
other $\rho $'s are smaller $\rho -\rho _{0}<0$ . Using the notation%
\begin{equation}
\rho -\rho _{0}=-\varepsilon <0  \label{eq38}
\end{equation}%
the denominator becomes $\frac{2a}{3d}\left[ -\varepsilon ^{3}+s\varepsilon
^{2}+t\varepsilon \right] $ where%
\begin{eqnarray}
s &\equiv &3\rho _{0}>0\ \ \ \ \ \ \ \ \ \ \ \ \ \ \ \ \ \ \ \ \ \ \ \ \left[
m^{-3}\right]  \label{eq39} \\
t &\equiv &\frac{3b}{a}I^{K}-3\rho _{0}^{2}>0\ \ \ \ \ \ \ \ \ \ \ \left[
m^{-6}\right]  \nonumber
\end{eqnarray}%
and the intagral%
\begin{equation}
\frac{-d\varepsilon }{\left[ -\varepsilon ^{3}+s\varepsilon
^{2}+t\varepsilon \right] ^{1/2}}=\pm \sqrt{\frac{2a}{3d}}dx  \label{eq40}
\end{equation}

\paragraph{Digression on the magnitudes of the parameters $s$ and $t$}

We want to underline a particularity of the problem connected with the
estimation of the orders of magnitude of the terms involved in these
equations. This problem will be found under different manifestations several
times below.

Estimation of the magnitude of the parameters $s$ and $t$,%
\begin{equation}
s\sim 3\times 10^{23}\ \left( m^{-3}\right)  \label{eq41}
\end{equation}%
\begin{equation}
t=\frac{3b}{a}I^{K}-3\rho _{0}^{2}\sim 10^{21}\left( \frac{1}{m^{6}}\right)
-3\times 10^{46}\ \left( \frac{1}{m^{6}}\right)  \label{eq42}
\end{equation}%
At the first sight $t$ is \emph{negative}, $t<0$ for $bI^{K}\sim 10^{8}\
\left( \frac{1}{m^{3}s}\right) $, where $I$ was taken $\sim 10^{15}\ \left( 
\frac{W}{m^{2}}\right) $. This is the uniform distribution in the cross
section of the beam and does not reflect the focusing effects, which can
lead to locally quasi-singular concentrations of $I$. We must take into
account that the first term can be much higher than it is here and this is
precisely the situation that is interesting for us. It will be much larger
when $bI^{K}$ will be multiplied by a coefficient \textquotedblleft $FACTOR$\textquotedblright. For the
following calculations we take 
\begin{equation}
t>0  \label{eq43}
\end{equation}%
which corresponds to the situation that the MPI is still higher than the
recombination.

\paragraph{Assuming $t>0$ MPI higher than recombination}

We make an approximation%
\begin{equation}
\frac{d\varepsilon }{\sqrt{\varepsilon \left( s\varepsilon +t\right) }}=\mp 
\sqrt{\frac{2a}{3d}}dx  \label{eq44}
\end{equation}%
by ignoring the high order $\varepsilon ^{3}$. Neglecting $\varepsilon ^{3}$
is equivalent to neglecting the highest effect of recombination. 
\begin{equation}
\frac{d\varepsilon }{\sqrt{\varepsilon \left( s\varepsilon +t\right) }}=%
\frac{1}{\sqrt{s}}\ln \left( \sqrt{s\left( s\varepsilon ^{2}+\varepsilon
t\right) }+2s\varepsilon +t\right) \ \ \textrm{for}\ \Delta <0\ \textrm{and}\
2s\varepsilon +t>\sqrt{-\Delta }=t  \label{eq45}
\end{equation}%
(Gradshtein Ryzhik 2.261). The equation%
\begin{equation}
\frac{d\varepsilon }{\sqrt{\varepsilon \left( s\varepsilon +t\right) }}=\mp 
\sqrt{\frac{2a}{3d}}dx  \label{eq46}
\end{equation}%
for $\varepsilon \equiv \rho _{0}-\rho \left( x\right) \geq 0$ is now
integrated%
\begin{equation}
\sqrt{\frac{s^{2}}{t^{2}}\varepsilon ^{2}+\frac{s}{t}\varepsilon }+2\frac{s}{%
t}\varepsilon +1=\exp \left[ \mp \sqrt{3\rho _{0}}\sqrt{\frac{2a}{3d}}\left(
x-x_{0}\right) \right]  \label{eq47}
\end{equation}%
where $x_{0}$ corresponds to the position where $\varepsilon =0$, which is
the same where the derivative of $\rho \left( x\right) $ is zero. Let%
\begin{equation}
y\equiv \frac{s}{t}\varepsilon  \label{eq49}
\end{equation}

\textbf{NOTE }regarding the magnitude and sign for the new variable $y$.

The magnitude is%
\begin{equation}
\left\vert y\right\vert =\left\vert \frac{3\times 10^{23}\ \left( \frac{1}{%
m^{3}}\right) }{10^{21}\left( \frac{1}{m^{6}}\right) -3\times 10^{46}\
\left( \frac{1}{m^{6}}\right) }\times 10^{23}\right\vert \sim 1\ 
\label{eq50}
\end{equation}

As results from $I\sim 10^{15}\ \left( \frac{W}{m^{2}}\right) $ the first
term in the expression of $t$ is much smaller than the second%
\begin{equation}
t=\frac{3b}{a}I^{K}-3\rho _{0}^{2}\sim 10^{21}\left( \frac{1}{m^{6}}\right)
-3\times 10^{46}\ \left( \frac{1}{m^{6}}\right)  \label{eq51}
\end{equation}%
and this would mean $y<0$. This has been discussed above. It is the
situation where we use the whole intensity of the beam without taking into
account the \emph{focalization} that is the origin of the formation of
stripes. Certainly we cannot assume that the focalization is quasi-singular,
with locally extremely high value for $I$ but we still must assume that the
formation of plasma (MPI $\sim bI^{K}$) is possible and the recombination
and diffusion just shape the profile.

Then 
\begin{equation}
t>0\ \ \ \textrm{and\ \ \ }t\sim 10^{46}\ \left( \frac{1}{m^{6}}\right)
\label{eq52}
\end{equation}%
It follows that%
\begin{equation}
y>0  \label{eq53}
\end{equation}%
\emph{\ }

\bigskip

We introduce the notation%
\begin{equation}
h\equiv \exp \left[ \mp \sqrt{3\rho \left( 0\right) }\sqrt{\frac{2a}{3d}}%
\left( x-x_{0}\right) \right]  \label{eq55}
\end{equation}%
and make few estimations. Since%
\begin{eqnarray}
\frac{2}{3}\frac{a}{d} &\sim &0.6\times \frac{5\times 10^{-13}}{10}\frac{%
\left( \frac{m^{3}}{s}\right) }{\left( \frac{m^{2}}{s}\right) }  \label{eq56}
\\
&\sim &3\times 10^{-14}\ \ \left( m\right)  \nonumber
\end{eqnarray}%
for $\rho \left( 0\right) \sim 10^{23}\ \left( m^{-3}\right) $. The
combination at the exponent%
\begin{equation}
\sqrt{3\rho \left( 0\right) }\sqrt{\frac{2a}{3d}}\approx 10^{5}\ \ \left( 
\frac{1}{m}\right)  \label{eq57}
\end{equation}

We find that $h$ verifies the necessary constraint $h\ll 1$. Introducing the
notation%
\begin{equation}
\sqrt{3\rho \left( 0\right) }\sqrt{\frac{2a}{3d}}\equiv \frac{1}{\xi }\
\left( \frac{1}{m}\right)  \label{eq58}
\end{equation}%
with units $\left[ \xi \right] =m$ we have%
\begin{equation}
h=\exp \left( -\frac{x-x_{0}}{\xi }\right)  \label{eq59}
\end{equation}

The equation becomes%
\begin{equation}
\sqrt{y^{2}+y}+2y+1=h  \label{eq60}
\end{equation}%
Returning to $\varepsilon $ we have%
\begin{equation}
\varepsilon =\frac{\frac{3b}{a}I^{K}-3\rho _{0}^{2}}{3\rho _{0}}\left[
0.1\pm 0.6\sqrt{1-h^{2}}\right]  \label{eq62}
\end{equation}%
\begin{equation}
\rho _{0}-\rho \left( x\right) =\left[ \frac{b}{a}\frac{I^{K}}{\rho _{0}}%
-\rho _{0}\right] \left[ 0.1\pm 0.6\sqrt{1-h^{2}}\right] >0  \label{eq63}
\end{equation}%
Note%
\begin{equation}
p\equiv \frac{b}{a}\frac{I^{K}}{\rho _{0}}-\rho _{0}  \label{eq64}
\end{equation}%
we have 
\begin{equation}
\rho \left( x\right) =\rho _{0}-p\frac{1}{10}\left( 5\mp 3h^{2}\right)
\label{eq67}
\end{equation}%
We argue that the sign $+$ must be chosen. This is because we want that the
overall term $-p\frac{1}{10}\left( 5+3h^{2}\right) $ to remain negative
since this reflects our choice of regime: fast generation of plasma through
ionization followed by diffusion and recombination still under a source
coming from MPI. If instead we had coosen $0.5\mp 0.3h^{2}=\frac{1}{10}%
\left( 5-3h^{2}\right) $ the term $-p\frac{1}{10}\left( 5-3h^{2}\right) $
were less negative.

\bigskip

\paragraph{NOTE on the magnitude of the parameter $p$}

The notation used above introduces%
\begin{equation}
p\equiv \frac{b}{a}\frac{I^{K}}{\rho _{0}}-\rho _{0}  \label{eq68}
\end{equation}%
As explained, the strong focalizaion that leads to plasma formation means
that the assumption $bI^{K}\sim 10^{8}\ \left( \frac{1}{m^{3}s}\right) $ is
an underestimation. The MPI term should generically be multiplied with a $%
FACTOR$ that represents the amplification in a spot that initiate a
filament. Then%
\begin{equation}
p\sim FACTOR\times 0.2\times 10^{-2}\left( \frac{1}{m^{3}}\right)
-10^{23}\left( \frac{1}{m^{3}}\right)  \label{eq69}
\end{equation}

For example, for an increase in the amplitude of electric field $E$ with a
factor of $100$, the amplification of the MPI term is $FACTOR=\left[ \left(
10^{2}\right) ^{2}\right] ^{K}=10^{28}$ for $K=7$ leading to%
\begin{equation}
p\sim 0.2\times 10^{-2}\times 10^{28}-10^{23}\ \left( \frac{1}{m^{3}}\right)
\label{eq72}
\end{equation}%
The parameter $p$ must be considered postive and with a magnitude similar to
the one of the two competing components, $p\sim 10^{23}$ $\left(
m^{-3}\right) $.

\bigskip

Finally we return to our equation%
\begin{equation}
\frac{\partial I}{\partial z}=D\frac{\partial ^{2}I}{\partial x^{2}}-\frac{%
\delta F}{\delta I}+\alpha I^{2}-\alpha ^{\prime }\rho I  \label{eq73}
\end{equation}%
where we replace%
\begin{equation}
\rho =\rho _{0}-\left( \frac{b}{a}\frac{I^{K}}{\rho _{0}}-\rho _{0}\right) 
\frac{1}{10}\left( 5+3h^{2}\right)  \label{eq74}
\end{equation}

Since we have assumed that the density that we study $\rho \left( x\right) $
is smaller (due to depletion by diffusion and recombination) than the
density created at the maximum of the focalization of $I$, which is the
maximum $\rho _{0}$, 
\begin{equation}
\rho \left( x\right) -\rho _{0}<0  \label{eq76}
\end{equation}%
$p$ must be positive such that the substraction to be correct%
\begin{equation}
\rho \left( x\right) =\rho _{0}-p\frac{1}{10}\left( 5+3h^{2}\right) <\rho
_{0}  \label{eq77}
\end{equation}

It is convenient to separate the expression of the density%
\begin{equation}
\rho \left( x\right) =-I^{K}w_{1}\left( x\right) +w_{2}\left( x\right)
\label{eq78}
\end{equation}%
\begin{eqnarray}
w_{1}\left( x\right) &\equiv &\frac{1}{10\rho _{0}}\frac{b}{a}\left(
5+3h^{2}\right)  \label{eq79} \\
w_{2}\left( x\right) &\equiv &\rho _{0}+\rho _{0}\frac{1}{10}\left(
5+3h^{2}\right)  \nonumber
\end{eqnarray}

The equation for $I$ becomes%
\begin{equation}
\frac{\partial I}{\partial z}=D\frac{\partial ^{2}I}{\partial x^{2}}-\frac{%
\delta F}{\delta I}+\alpha I^{2}-\alpha ^{\prime }I\left[ -I^{K}w_{1}\left(
x\right) +w_{2}\left( x\right) \right]  \label{eq81}
\end{equation}

\section{The stabilization of the stripe}

We start from the differential equations for the \emph{activator} field (the
intensity $I$).

The equation 
\begin{equation}
\frac{\partial I}{\partial z}=D\frac{\partial ^{2}I}{\partial x^{2}}-\frac{%
\delta F}{\delta I}+\alpha I^{2}-\alpha ^{\prime }\rho I  \label{eq82}
\end{equation}

It can be derived from%
\begin{equation}
\mathcal{W}_{I}=\int dx\left[ \frac{1}{2}D\left( \frac{\partial I}{\partial x%
}\right) ^{2}+F\left[ I\right] -\alpha \frac{I^{3}}{3}\right] +\alpha
^{\prime }\frac{1}{2}\int dx\rho \left( x\right) I^{2}\left( x\right)
\label{eq83}
\end{equation}

And, the equation for the density $\rho $ is%
\begin{equation}
\frac{\partial \rho }{\partial t}=d\frac{\partial ^{2}\rho }{\partial x^{2}}%
-a\rho ^{2}+bI^{K}  \label{eq84}
\end{equation}%
with the Energy functional%
\begin{equation}
\mathcal{W}_{\rho }=\int dx\left[ \frac{1}{2}d\left( \frac{\partial \rho }{%
\partial x}\right) ^{2}+a\frac{\rho ^{3}}{3}\right] +b\int dx\rho I^{K}
\label{eq85}
\end{equation}

We follow the work by Goldstein  \cite{goldsteinlabyrinth} to study
the evolution of a stripe $I=I_{\max }$ between regions (also stripes) of $%
I=0$.

\subsection{The variational equations}

\subsubsection{Variational equation for the intensity}

The equation for $I$ can be written in variational form. We separate the
non-coupled parts in the functionals%
\begin{equation}
\mathcal{W}_{I}=\mathcal{E}_{I}+\mathcal{F}_{I}  \label{eq86}
\end{equation}%
\begin{equation}
\mathcal{E}_{I}=\int dx\left[ \frac{1}{2}D\left( \frac{\partial I}{\partial x%
}\right) ^{2}+F\left[ I\right] -\alpha \frac{I^{3}}{3}\right]  \label{eq87}
\end{equation}%
and the coupled part%
\begin{equation}
\mathcal{F}_{I}\mathcal{=}\alpha ^{\prime }\int dx\frac{1}{2}\rho \left(
x\right) I^{2}\left( x\right)  \label{eq89}
\end{equation}%
and calculate first for $I$. After an integration by parts 
\begin{equation}
\mathcal{E}_{I}\left[ I\right] =\int dx\left[ -\frac{1}{2}DI\left( \frac{%
\partial ^{2}I}{\partial x^{2}}\right) +F\left[ I\right] -\alpha \frac{I^{3}%
}{3}\right]  \label{eq90}
\end{equation}%
By functional integration of $\mathcal{E}_{I}$ to $I\left( x\right) $ we get
a $\delta \left( x-x^{\prime }\right) $ factor which will be integrated over 
$x^{\prime }$ and selects precisely the terms calculated at $x$, \emph{i.e. }%
\ the equation. The integration of product of identical functions like $%
\left( \partial I/\partial x\right) $ will occur twice 
\begin{equation}
\frac{\delta \mathcal{E}_{I}}{\delta I}=-D\left( \frac{\partial ^{2}I}{%
\partial x^{2}}\right) +\frac{\delta F}{\delta I}-\alpha I^{2}  \label{eq91}
\end{equation}%
To this equation we add the result of functional variation of the coupling
term%
\begin{equation}
\frac{\delta \mathcal{F}_{I}}{\delta I}=\alpha ^{\prime }\rho \left(
x\right) I\left( x\right)  \label{eq92}
\end{equation}%
\begin{equation}
\frac{\delta \mathcal{F}_{I}}{\delta \rho }=\alpha ^{\prime }I^{2}\left(
x\right)  \label{eq93}
\end{equation}

The equation for the variable $I$ is%
\begin{equation}
\frac{\partial I}{\partial z}=D\frac{\partial ^{2}I}{\partial x^{2}}-\frac{%
\delta F}{\delta I}+\alpha I^{2}-\alpha ^{\prime }\rho I  \label{eq94}
\end{equation}%
can now be written 
\begin{equation}
\frac{\partial I}{\partial z}=-\frac{\delta \mathcal{E}_{I}}{\delta I}-\frac{%
\delta \mathcal{F}_{I}}{\delta I}  \label{eq101}
\end{equation}

\subsubsection{Variational equation for the density $\protect\rho $}

In an analogous calculation we separate in the energy functional the
coupling term 
\begin{equation}
\mathcal{W}_{\rho }=\mathcal{E}_{\rho }+\mathcal{F}_{\rho }  \label{eq102}
\end{equation}%
\begin{equation}
\mathcal{E}_{\rho }=\int dx\left[ \frac{1}{2}d\left( \frac{\partial \rho }{%
\partial x}\right) ^{2}+a\frac{\rho ^{3}}{3}\right]  \label{eq103}
\end{equation}%
\begin{equation}
\mathcal{F}_{\rho }\mathcal{=}b\int dx\frac{1}{2}\rho ^{2}\left( x\right)
I^{K}\left( x\right)  \label{eq104}
\end{equation}

Preparing for functional variation%
\begin{equation}
\mathcal{E}_{\rho }\left[ \rho \right] =\int dx\left[ -\frac{1}{2}d\ \rho
\left( \frac{\partial ^{2}\rho }{\partial x^{2}}\right) +a\frac{\rho ^{3}}{3}%
\right]  \label{eq105}
\end{equation}%
\begin{equation}
\frac{\delta \mathcal{E}_{\rho }}{\delta \rho }=-d\left( \frac{\partial
^{2}\rho }{\partial x^{2}}\right) +a\rho ^{2}  \label{eq106}
\end{equation}%
\begin{equation}
\frac{\delta \mathcal{F}_{\rho }}{\delta I}=bK\rho \left( x\right)
I^{K-1}\left( x\right)  \label{eq108}
\end{equation}%
\begin{equation}
\frac{\delta \mathcal{F}_{\rho }}{\delta \rho }=bI^{K}\left( x\right)
\label{eq109}
\end{equation}%
The equation of motion%
\begin{equation}
\frac{\partial \rho }{\partial t}=d\frac{\partial ^{2}\rho }{\partial x^{2}}%
-a\rho ^{2}+bI^{K}  \label{eq110}
\end{equation}%
is written as%
\begin{equation}
\frac{\partial \rho }{\partial t}=-\frac{\delta \mathcal{E}_{\rho }}{\delta
\rho }+\frac{\delta \mathcal{F}_{\rho }}{\delta \rho }  \label{eq111}
\end{equation}

\bigskip

\subsection{Is-there a gradient flow?}

An important factor in the formation of a \emph{labyrinth} pattern for an 
\emph{activator-inhibitor} system is reduction of the dynamics to the \emph{%
gradient flow} \cite{desaikapral}, \cite{goldsteinlabyrinth}. We would like
to check that the same structure exists for the two fields $\left( I,\rho
\right) $. We take infinitely fast inhibitor%
\begin{equation}
\frac{\partial \rho }{\partial t}=-\frac{\delta \mathcal{E}_{\rho }}{\delta
\rho }+\frac{\delta \mathcal{F}_{\rho }}{\delta \rho }=0  \label{eq120}
\end{equation}%
and calculate $\frac{\partial }{\partial z}\left( \mathcal{E}_{I}+\mathcal{F}%
_{I}\right) $. We use Eqs.(\ref{eq86}) - (\ref{eq89})%
\begin{eqnarray}
\frac{\partial \mathcal{E}_{I}}{\partial z} &=&\frac{\partial \mathcal{E}_{I}%
}{\partial I}\frac{\partial I}{\partial z}+\frac{\partial \mathcal{E}_{I}}{%
\partial \rho }c\frac{\partial \rho }{\partial z}  \label{eq121} \\
&=&\int dx\left[ -D\left( \frac{\partial ^{2}I}{\partial x^{2}}\right) +%
\frac{\delta F}{\delta I}-\alpha I^{2}\right] \times \left[ D\frac{\partial
^{2}I}{\partial x^{2}}-\frac{\delta F}{\delta I}+\alpha I^{2}-\alpha
^{\prime }\rho I\right]  \nonumber
\end{eqnarray}%
In the first square paranthesis we add and substract what is missing for the
expression inside to become $-\frac{\partial I}{\partial z}$ which means the
second square paranthesis with negative sign%
\begin{eqnarray}
\frac{\partial \mathcal{E}_{I}}{\partial z} &=&\frac{\partial \mathcal{E}_{I}%
}{\partial I}\frac{\partial I}{\partial z}=\int dx\left\{ -\left[ D\frac{%
\partial ^{2}I}{\partial x^{2}}-\frac{\delta F}{\delta I}+\alpha
I^{2}-\alpha ^{\prime }\rho I\right] ^{2}\right.  \label{eq123} \\
&&\ \ \ \ \ \ \ \ \ \ \ \ \ \ \left. -\alpha ^{\prime }\rho \left( x\right)
I\left( x\right) \left[ D\frac{\partial ^{2}I}{\partial x^{2}}-\frac{\delta F%
}{\delta I}+\alpha I^{2}-\alpha ^{\prime }\rho I\right] \right\}  \nonumber
\end{eqnarray}%
For the second part we have%
\begin{eqnarray}
\frac{\partial \mathcal{F}_{I}}{\partial z} &=&\frac{\partial \mathcal{F}_{I}%
}{\partial z}=\frac{\delta \mathcal{F}_{I}}{\delta I}\frac{\partial I}{%
\partial z}+\frac{\delta \mathcal{F}_{I}}{\delta \rho }c\frac{\partial \rho 
}{\partial t}=\frac{\delta \mathcal{F}_{I}}{\delta I}\frac{\partial I}{%
\partial z}  \label{eq124} \\
&=&\int dx\left[ \alpha ^{\prime }\rho \left( x\right) I\left( x\right) %
\right] \times \left[ D\frac{\partial ^{2}I}{\partial x^{2}}-\frac{\delta F}{%
\delta I}+\alpha I^{2}-\alpha ^{\prime }\rho I\right]  \nonumber
\end{eqnarray}%
Adding the two expressions we obtain%
\begin{equation}
\frac{\partial }{\partial z}\left( \mathcal{E}_{I}+\mathcal{F}_{I}\right)
=-\int dx\left[ D\frac{\partial ^{2}I}{\partial x^{2}}-\frac{\delta F}{%
\delta I}+\alpha I^{2}-\alpha ^{\prime }\rho I\right] ^{2}=-\int dx\left( 
\frac{\partial I}{\partial z}\right) ^{2}<0  \label{eq125}
\end{equation}%
and this confirms that we have a \emph{gradient flow}.

The fact that the evolution of the intensity $I$ is a gradient flow supports
the idea that the optical turbulence and the activator-inhibitor have the
same mathematical nature.

\bigskip

\subsection{The energy of a stripe}

We consider a stripe belonging to the cluster of high intensity, of
time-dependent width $2Q$, $x\in \left[ -Q,+Q\right] $ . The axis of the
stripe is considered a line and does not intervene in the calculation below.
The energy functional for $I$ will be used to calculate the energy of the
stripe on unit length along the axis 
\begin{equation}
\mathcal{W}_{I}=\int dx\left[ \frac{1}{2}D\left( \frac{\partial I}{\partial x%
}\right) ^{2}+F\left[ I\right] -\alpha \frac{I^{3}}{3}\right] +\alpha
^{\prime }\frac{1}{2}\int dx\rho \left( x\right) I^{2}\left( x\right)
\label{eq126}
\end{equation}

According to the method developed by Goldstein \cite%
{goldsteinlabyrinth}, we must evaluate the contributions.

(1) The \textquotedblleft line tension\textquotedblright\ arises from the
gradient at the front (interface) 
\begin{equation}
\gamma \sim \int_{front}dxD\left( \frac{\partial I}{\partial x}\right) ^{2}
\label{eq127}
\end{equation}

(2) The pressure $\Pi $ \ is the density of the energy contained in the
stripe relative to the \textquotedblleft empty\textquotedblright\ regions
around 
\begin{equation}
\Pi =\frac{1}{2Q}\int_{-Q}^{Q}dx\left\{ F\left[ I=I_{\max }\right] -F\left[
I=0\right] \right\} =\Delta F  \label{eq128}
\end{equation}

(3) For the third term we have to introduce the expression of $\rho \left(
x\right) $ that we have calculated.

We remember that the stripe is defined by $I=I_{\max }=$ const on a spatial
region of length $2Q$ bounded by zones \textquotedblleft
empty\textquotedblright\ of intensity, $I=0$. 
\begin{equation}
I=I_{\max }\left[ \Theta \left( x+Q\right) -\Theta \left( x-Q\right) \right]
\label{eq129}
\end{equation}%
that when we integrate over the stripe where $I=I_{\max }=$const we have%
\begin{equation}
\int_{-Q}^{Q}dx\left[ -\alpha \frac{I_{\max }^{3}}{3}\right] =-\alpha \frac{%
I_{\max }^{3}}{3}2Q  \label{eq130}
\end{equation}%
and will contribute to variational terms.

The last term in the expression of $\mathcal{W}_{I}$ comes from the coupling
with $\rho \left( x\right) $ 
\begin{equation}
\int_{-Q}^{Q}dx\left[ \alpha ^{\prime }\frac{1}{2}\rho \left( x\right)
I^{2}\left( x\right) \right] =\frac{\alpha ^{\prime }}{2}I_{\max
}^{2}\int_{-Q}^{Q}dx\rho \left( x\right)  \label{eq131}
\end{equation}%
where $\rho \left( x\right) $ is given in terms of $h\left( x\right) $.

The total energy%
\begin{equation}
\Delta E\left( Q\right) \sim 2\gamma +2Q\Delta F-\alpha \frac{I_{\max }^{3}}{%
3}2Q+\frac{\alpha ^{\prime }}{2}I_{\max }^{2}\int_{-Q}^{Q}dx\rho \left(
x\right)  \label{eq134}
\end{equation}%
is the Lagrangian density for time-independent state%
\begin{equation}
\mathcal{L}\left[ I\right] =-\Delta E\left( Q\right)  \label{eq135}
\end{equation}

The expression of $\mathcal{L}$ must be employed in the Euler Lagrange
variational equation. However there is an additional term that should be
considered, \emph{i.e.} a dissipative term \cite{goldsteinlabyrinth}%
\begin{equation}
\mathcal{R}\left[ \frac{\partial I}{\partial z}\right] =\int_{-\infty
}^{\infty }dx\frac{1}{2}\left( \frac{\partial I}{\partial z}\right) ^{2}
\label{eq138}
\end{equation}%
and the Euler-Lagrange equation is%
\begin{equation}
\frac{d}{dz}\frac{\delta \mathcal{L}}{\delta \left( \frac{\partial I}{%
\partial z}\right) }-\frac{\delta \mathcal{L}}{\delta I}=-\frac{\delta 
\mathcal{R}}{\delta \left( \frac{\partial I}{\partial z}\right) }
\label{eq139}
\end{equation}%
The functional that is considered dissipative, $\mathcal{R}$, will be
calculated replacing 
\begin{equation}
\frac{\partial I}{\partial z}=\frac{\partial I}{\partial x}\frac{\partial x}{%
\partial z}  \label{eq140}
\end{equation}%
and taking into account that there is the boundary condition which is time
dependent, \emph{i.e.} $x\left( t\right) =Q\left( t\right) $. 
\begin{equation}
\mathcal{R}\left[ \frac{\partial I}{\partial z}\right] =\left( \frac{%
\partial Q}{\partial z}\right) ^{2}\int_{front}dx\left( \frac{\partial I}{%
\partial x}\right) ^{2}  \label{eq143}
\end{equation}%
The integral involves the \textquotedblleft line tension\textquotedblright\
and is replaced by 
\begin{equation}
\mathcal{R}\left[ \frac{\partial I}{\partial z}\right] \sim \left( \frac{%
\partial Q}{\partial z}\right) ^{2}\frac{\gamma }{D}  \label{eq146}
\end{equation}%
and%
\begin{eqnarray}
\frac{\delta \mathcal{R}}{\delta \left( \frac{\partial I}{\partial z}\right) 
} &=&\frac{\delta }{\delta \left( \frac{\partial I}{\partial z}\right) }%
\left\{ \left( \frac{\partial Q}{\partial z}\right) ^{2}\frac{\gamma }{D}%
\right\}  \label{eq147} \\
&=&\frac{\partial Q}{\partial z}\frac{2\gamma }{D}  \nonumber
\end{eqnarray}

the variational equation becomes%
\begin{eqnarray}
\mathcal{L}\left[ I\right] &=&-\Delta E\left( Q\right)  \label{eq148} \\
&=&-\left[ 2\gamma +2Q\Delta F-\alpha \frac{I_{\max }^{3}}{3}2Q+\frac{\alpha
^{\prime }}{2}I_{\max }^{2}\int_{-Q}^{Q}dx\rho \left( x\right) \right] 
\nonumber
\end{eqnarray}%
or%
\begin{equation}
\frac{2\gamma }{D}\frac{\partial Q}{\partial z}=-\frac{\partial \left(
\Delta E\right) }{\partial Q}  \label{eq149}
\end{equation}

It results%
\begin{equation}
\frac{\gamma }{D}\frac{\partial Q}{\partial z}=-\left[ \Delta F-\alpha \frac{%
I_{\max }^{3}}{3}+\frac{\alpha ^{\prime }}{2}I_{\max }^{2}\frac{1}{2}\frac{%
\partial }{\partial Q}\int_{-Q}^{Q}dx\rho \left( x\right) \right]
\label{eq150}
\end{equation}

To advance we have to examine the last term. It has been derived above, Eq.(%
\ref{eq78}), the following expression for the density of plasma electrons
determined by : the intensity $I$, the recombination and diffusion%
\begin{equation}
\rho \left( x\right) =-I^{K}w_{1}\left( x\right) +w_{2}\left( x\right)
\label{eq151}
\end{equation}%
Now we make more explicit the last term%
\begin{eqnarray}
&&\frac{\alpha ^{\prime }}{2}I_{\max }^{2}\frac{1}{2}\frac{\partial }{%
\partial Q}\int_{-Q}^{Q}dx\rho \left( x\right)  \label{eq152} \\
&=&\frac{\alpha ^{\prime }}{2}I_{\max }^{2}\frac{1}{2}\frac{\partial }{%
\partial Q}\left[ \int_{-Q}^{Q}dxw_{2}\left( x\right) -I_{\max
}^{K}\int_{-Q}^{Q}dxw_{1}\left( x\right) \right]  \nonumber
\end{eqnarray}

We have%
\begin{equation}
\int_{-Q}^{Q}dxw_{2}\left( x\right) =2Q\rho _{0}\frac{3}{2}+\rho _{0}\frac{3%
}{10}\xi \exp \left( \frac{2x_{0}}{\xi }\right) \sinh \left( \frac{Q}{\xi /2}%
\right)  \label{eq154}
\end{equation}%
and%
\begin{equation}
\int_{-Q}^{Q}dxw_{1}\left( x\right) =\frac{b}{a}\frac{Q}{\rho _{0}}+\frac{3b%
}{10a\rho _{0}}\xi \exp \left( \frac{2x_{0}}{\xi }\right) \sinh \left( \frac{%
Q}{\xi /2}\right)  \label{eq156}
\end{equation}%
Replacing%
\begin{eqnarray}
&&\frac{\partial }{\partial Q}\int_{-Q}^{Q}dx\rho \left( x\right) =-I^{K}%
\frac{b}{a\rho _{0}}+3\rho _{0}  \label{eq157} \\
&&+\frac{3}{5}\left[ -I^{K}\frac{b}{a\rho _{0}}+\rho _{0}\right] \exp \left( 
\frac{2x_{0}}{\xi }\right) \cosh \left( \frac{Q}{\xi /2}\right)  \nonumber
\end{eqnarray}%
we introduce the notation%
\begin{eqnarray}
q &\equiv &-I^{K}\frac{b}{a\rho _{0}}+3\rho _{0}  \label{eq158} \\
&=&-p+2\rho _{0}  \nonumber
\end{eqnarray}%
and the result is represented as%
\begin{equation}
\frac{\partial }{\partial Q}\int_{-Q}^{Q}dx\rho \left( x\right) =q-p\frac{3}{%
5}\exp \left( \frac{2x_{0}}{\xi }\right) \cosh \left( \frac{Q}{\xi /2}\right)
\label{eq159}
\end{equation}

\bigskip

We can now write the functional%
\begin{eqnarray}
\frac{\gamma }{D}\frac{\partial Q}{\partial z} &=&-\Delta F+\alpha \frac{%
I_{\max }^{3}}{3}-\alpha ^{\prime }\frac{I_{\max }^{2}}{4}q  \label{eq160} \\
&&+\alpha ^{\prime }\frac{I_{\max }^{2}}{4}p\frac{3}{5}\exp \left( \frac{%
2x_{0}}{\xi }\right) \cosh \left( \frac{Q}{\xi /2}\right)  \nonumber
\end{eqnarray}

A stationary state for the stripe exists when $\partial Q/\partial z=0$,
which has the approximative form%
\begin{equation}
6\left( \frac{\Delta F}{I_{\max }^{2}\alpha ^{\prime }p}-\frac{\alpha }{%
\alpha ^{\prime }p}+\frac{q}{p}\right) \exp \left( \frac{2x_{0}}{\xi }%
\right) =\cosh \left( \frac{Q}{\xi /2}\right)  \label{eq162}
\end{equation}%
We introduce the notation%
\begin{equation}
r\equiv 6\left( \frac{\Delta F}{pI_{\max }^{2}\alpha ^{\prime }}-\frac{%
\alpha }{p\alpha ^{\prime }}I_{\max }-1+\frac{2\rho _{0}}{p}\right)
\label{eq163}
\end{equation}%
and for a stabilization of the stripe width we need $r>1$. \ For an
evaluation we use the magnitudes chosen above and adopt a hypothesis on the
difference between the potential energies of the two basic states 
\begin{equation}
\frac{\Delta F}{pI_{\max }^{2}\alpha ^{\prime }}\sim 1  \label{eq165}
\end{equation}%
We conclude that the terms in $r$ can lead to a negative value which means
that there is no stabilization of the stripes.

If however the concentration of beam energy renders $I_{\max }$ higher by
orders of magnitude compared with the uniformly distributed input $I$ then $%
r $ can be positive and of order few units. In this case, adopting $x_{0}=0$%
, we solve $u^{2}-2ru+1=0$ and find $u=\exp \left( \frac{2Q}{\xi }\right) $.
Then $2Q\sim \xi \ln r$ leads to a rough estimation%
\begin{equation}
Q\gtrsim \xi \sim 10^{-5}\ \left( m\right)  \label{eq166}
\end{equation}%
where we used the estimation%
\begin{eqnarray}
\xi &\equiv &\left[ \sqrt{3\rho _{0}}\sqrt{\frac{2a}{3d}}\right] ^{-1}
\label{eq169} \\
&\sim &10^{-5}\ \left( m\right)  \nonumber
\end{eqnarray}

The result is smaller than the width that can be retrived from the pictures
obtained experimentally by Ettoumi \emph{et al}. \cite{phasetransition}
where one can infer an average width $\sim 10^{-4}\ \left( m\right) $.

We can improve the analytical framework with the purpose of a better
description of the balance between numbers of very high magnitude ($\sim
10^{23}$ ) that are substracted in the competition between Kerr and plasma
nonlinearities. We will need new technical methods and some numerical work
in parallel.

\bigskip

\section{Conclusion}

The previous work \cite{florinmadilabyrinth} has advanced a hypothesis that
there is a common mathematical structure underlying the optical turbulence
and the gradient flow of some nonlinear reaction diffusion system. The
common ground is the activator-inhibitor dynamics where two fields, one
auto-catalitic and the other acting to limit and inhibit the expansion of
the first, compete and generate a complicated pattern. The distribution of
the intensity of the laser pulse is mainly the result of self-focusing
(Kerr) nonlinearity and defocusing effect of the plasma created by
ionization. The basic model of self-focalization is excatly integrable and
we argue that starting from here one can construct a mathematical model that
incorporates the known physical processes of beam propagation in a way that
makes transparent the analogy with the activator-inhibitor dynamics. The
constructed model yields the analytical form Eq.(\ref{eq18}) which, together with
the equation for the density $\rho \left( x\right) $ indeed shows the
dynamics of activator-inhibitor type.

We show that it has the structure of gradient flow and we study the possible
regimes consisting of suppression or, alternatively, saturation to a finite
width of the stripe belonging to the cluster of high intensity.

As explained in the previous work, there is a practical utility in revealing
this parallel between optical turbulence and the activator-inhibitor
dynamics. The latter has been thoroughly investigated and many aspects can
now be mapped on the corresponding behavior of the intensity in the
transversal plane of a laser beam: formation of spots of high intensity,
possibly with crystal spatial distribution, etc.

A numerical study devoted to this analogue mathematical behavior may be
useful.

\bigskip
{\bf Aknowledgment}  This work has been supported in part by the Contract 4N/2016 of the Project PN 16 47 01 01 of the Romanian Minsitry of Education and Scientific Research.

\begin{appendices}

\section{Appendix A. The hodograph transformation}

\label{App:AppendixA}

\renewcommand{\theequation}{A.\arabic{equation}} \setcounter{equation}{0}

We adopt the standard treatment of Trubnikov and Zhdanov \cite%
{trubnikovzhdanov} of the nonlinear self-focusing. See also Appendix A of
Ref.\cite{florinmadireversalrotation}. 

The equations are%
\begin{eqnarray}
\frac{\partial I}{\partial z}+\frac{\partial }{\partial x}\left( vI\right) 
&=&0  \label{a1} \\
\frac{\partial v}{\partial z}+v\frac{\partial v}{\partial x} &=&c_{0}\frac{%
\partial }{\partial x}\left( \frac{I}{I_{0}}\right)   \nonumber
\end{eqnarray}%
where%
\begin{eqnarray}
v &=&\frac{\partial S}{\partial x}  \label{a2} \\
&=&\textrm{transversal derivative of the eikonal}  \nonumber
\end{eqnarray}%
\begin{equation}
c_{0}^{2}=\frac{\varepsilon _{2}}{2\varepsilon _{0}}I_{0}  \label{a3}
\end{equation}%
\begin{equation}
I=a^{2}  \label{a4}
\end{equation}%
\begin{eqnarray}
I_{0} &=&a_{0}^{2}  \label{a5} \\
&=&\textrm{intensity at the entrance in the medium}  \nonumber
\end{eqnarray}%
and%
\begin{equation}
A\left( z,x\right) =a\left( z,x\right) \exp \left[ ikS\left( z,x\right) %
\right]   \label{a6}
\end{equation}

The variables are%
\begin{equation}
\left( x,z\right) \rightarrow \left( I,v\right)   \label{a7}
\end{equation}%
Now we apply the hodograph transformation to express $\left( x,z\right) $ in
terms of $\left( I,v\right) $ following closely the original treatment \cite%
{trubnikovbook}%
\begin{eqnarray}
\frac{dz}{dz} &=&1=\frac{\partial z}{\partial I}\frac{\partial I}{\partial z}%
+\frac{\partial z}{\partial v}\frac{\partial v}{\partial z}  \label{a8} \\
\frac{dz}{dx} &=&0=\frac{\partial z}{\partial I}\frac{\partial I}{\partial x}%
+\frac{\partial z}{\partial v}\frac{\partial v}{\partial x}  \nonumber
\end{eqnarray}%
and%
\begin{eqnarray}
\frac{dx}{dz} &=&0=\frac{\partial x}{\partial I}\frac{\partial I}{\partial z}%
+\frac{\partial x}{\partial v}\frac{\partial v}{\partial z}  \label{a9} \\
\frac{dx}{dx} &=&1=\frac{\partial x}{\partial I}\frac{\partial I}{\partial x}%
+\frac{\partial x}{\partial v}\frac{\partial v}{\partial x}  \nonumber
\end{eqnarray}%
This is a linear system with four equations and four unknowns. The first
equation from the first group and the first equation from the second group
are solved using the Jacobian%
\begin{eqnarray}
\det \left( 
\begin{array}{cc}
\frac{\partial z}{\partial I} & \frac{\partial z}{\partial v} \\ 
\frac{\partial x}{\partial I} & \frac{\partial x}{\partial v}%
\end{array}%
\right)  &=&\frac{\partial z}{\partial I}\frac{\partial x}{\partial v}-\frac{%
\partial z}{\partial v}\frac{\partial x}{\partial I}  \label{a10} \\
&=&-J  \nonumber
\end{eqnarray}%
Then%
\begin{equation}
\frac{\partial I}{\partial z}=-\frac{1}{J}\frac{\partial x}{\partial v}
\label{a11}
\end{equation}%
\begin{equation}
\frac{\partial v}{\partial z}=\frac{-1}{J}\left( -\frac{\partial x}{\partial
I}\right)   \label{a12}
\end{equation}

Now we repeat for: the second equation from the first group and the second
equation from the second group%
\begin{eqnarray}
0 &=&\frac{\partial z}{\partial I}\frac{\partial I}{\partial x}+\frac{%
\partial z}{\partial v}\frac{\partial v}{\partial x}  \label{a13} \\
1 &=&\frac{\partial x}{\partial I}\frac{\partial I}{\partial x}+\frac{%
\partial x}{\partial v}\frac{\partial v}{\partial x}  \nonumber
\end{eqnarray}%
The result%
\begin{eqnarray}
\frac{\partial I}{\partial x} &=&-\frac{1}{J}\left( -\frac{\partial z}{%
\partial v}\right)  \label{14} \\
\frac{\partial v}{\partial x} &=&-\frac{1}{J}\frac{\partial z}{\partial I} 
\nonumber
\end{eqnarray}%
The result is%
\begin{eqnarray}
\frac{\partial v}{\partial z} &=&\frac{1}{J}\frac{\partial x}{\partial I}
\label{a16} \\
\frac{\partial I}{\partial z} &=&-\frac{1}{J}\frac{\partial x}{\partial v} 
\nonumber \\
\frac{\partial v}{\partial x} &=&-\frac{1}{J}\frac{\partial z}{\partial I} 
\nonumber \\
\frac{\partial I}{\partial x} &=&\frac{1}{J}\frac{\partial z}{\partial v} 
\nonumber
\end{eqnarray}

It is the time to replace these expressions in the \emph{Chaplygin }%
equations for self-focusing%
\begin{eqnarray}
\frac{\partial I}{\partial z}+\frac{\partial }{\partial x}\left( vI\right)
&=&0  \label{a17} \\
\frac{\partial v}{\partial z}+v\frac{\partial v}{\partial x} &=&c_{0}^{2}%
\frac{\partial }{\partial x}\left( \frac{I}{I_{0}}\right)  \nonumber
\end{eqnarray}%
where we carry out the derivations%
\begin{eqnarray}
\frac{\partial I}{\partial z}+\frac{\partial v}{\partial x}I+v\frac{\partial
I}{\partial x} &=&0  \label{a18} \\
\frac{\partial v}{\partial z}+v\frac{\partial v}{\partial x} &=&c_{0}^{2}%
\frac{\partial I}{\partial x}\frac{1}{I_{0}}  \nonumber
\end{eqnarray}%
and replace in the first equation%
\begin{equation}
-\frac{1}{J}\frac{\partial x}{\partial v}-\frac{1}{J}\frac{\partial z}{%
\partial I}I+v\left( \frac{1}{J}\frac{\partial z}{\partial v}\right) =0
\label{a19}
\end{equation}%
\begin{eqnarray}
\frac{\partial x}{\partial v}+\frac{\partial z}{\partial I}I-v\frac{\partial
z}{\partial v} &=&0  \label{a20} \\
&&\textrm{or}  \nonumber \\
\frac{\partial x}{\partial v} &=&v\frac{\partial z}{\partial v}-I\frac{%
\partial z}{\partial I}  \nonumber
\end{eqnarray}

Now we replace in the second equation%
\begin{equation}
\frac{\partial v}{\partial z}+v\frac{\partial v}{\partial x}=c_{0}^{2}\frac{%
\partial I}{\partial x}\frac{1}{I_{0}}  \label{a21}
\end{equation}%
it is%
\begin{eqnarray}
\frac{1}{J}\frac{\partial x}{\partial I}+v\left( -\frac{1}{J}\frac{\partial z%
}{\partial I}\right)  &=&\frac{c_{0}^{2}}{I_{0}}\frac{1}{J}\frac{\partial z}{%
\partial v}  \label{a22} \\
&&\textrm{or}  \nonumber \\
\frac{\partial x}{\partial I} &=&\frac{c_{0}^{2}}{I_{0}}\frac{\partial z}{%
\partial v}+v\frac{\partial z}{\partial I}  \nonumber
\end{eqnarray}%
We must take care of the mixed derivatives%
\begin{equation}
\frac{\partial ^{2}x}{\partial I\partial v}=\frac{\partial ^{2}x}{\partial
v\partial I}  \label{a23}
\end{equation}%
\begin{equation}
\frac{\partial }{\partial I}\left[ v\frac{\partial z}{\partial v}-I\frac{%
\partial z}{\partial I}\right] =\frac{\partial }{\partial v}\left[ \frac{%
c_{0}^{2}}{I_{0}}\frac{\partial z}{\partial v}+v\frac{\partial z}{\partial I}%
\right]   \label{a24}
\end{equation}%
From this%
\begin{equation}
\frac{\partial v}{\partial I}\frac{\partial z}{\partial v}+v\frac{\partial
^{2}z}{\partial I\partial v}-\frac{\partial z}{\partial I}-I\frac{\partial
^{2}z}{\partial I^{2}}=\frac{c_{0}^{2}}{I_{0}}\frac{\partial ^{2}z}{\partial
v^{2}}+\frac{\partial z}{\partial I}+v\frac{\partial ^{2}z}{\partial
v\partial I}  \label{a40}
\end{equation}%
We note that the second term from the LHS is reduced with the last term of
the RHS and that the first term in the LHS is identically zero since $v$ and 
$I$ are independent variables of the second set, just like $\left(
z,x\right) $.%
\begin{equation}
\frac{c_{0}^{2}}{I_{0}}\frac{\partial ^{2}z}{\partial v^{2}}+I\frac{\partial
^{2}z}{\partial I^{2}}+2\frac{\partial z}{\partial I}=0  \label{a41}
\end{equation}%
\begin{equation}
\frac{1}{I}\frac{\partial }{\partial I}\left( I^{2}\frac{\partial z}{%
\partial I}\right) +\frac{c_{0}^{2}}{I_{0}}\frac{\partial ^{2}z}{\partial
v^{2}}=0  \label{a42}
\end{equation}

We make the substitution%
\begin{eqnarray}
r &=&I^{2}  \label{a43} \\
s &=&\frac{1}{2}\frac{v}{\sqrt{c_{0}^{2}/I_{0}}}  \nonumber
\end{eqnarray}%
We calculate%
\begin{eqnarray}
I &=&r^{2}  \label{a44} \\
\frac{\partial }{\partial r} &=&\frac{\partial }{\partial I}\frac{\partial I%
}{\partial r}=2r\frac{\partial }{\partial I}  \nonumber \\
\frac{\partial }{\partial I} &=&\frac{1}{2r}\frac{\partial }{\partial r} 
\nonumber
\end{eqnarray}%
\begin{eqnarray}
s &=&v\frac{1}{2}\frac{1}{\sqrt{c_{0}^{2}/I_{0}}}  \label{a45} \\
v &=&\alpha s\ \ \ \textrm{where}\ \ \alpha \equiv 2\sqrt{c_{0}^{2}/I_{0}} 
\nonumber \\
\frac{\partial }{\partial s} &=&\frac{\partial v}{\partial s}\frac{\partial 
}{\partial v}=\alpha \frac{\partial }{\partial v}  \nonumber \\
\frac{\partial }{\partial v} &=&\frac{1}{\alpha }\frac{\partial }{\partial s}
\nonumber
\end{eqnarray}%
\begin{eqnarray}
\frac{1}{I}\frac{\partial }{\partial I}\left( I^{2}\frac{\partial z}{%
\partial I}\right) +\frac{c_{0}^{2}}{I_{0}}\frac{\partial ^{2}z}{\partial
v^{2}} &=&0  \label{a47} \\
\frac{1}{r^{2}}\frac{1}{2r}\frac{\partial }{\partial r}\left( r^{4}\frac{1}{%
2r}\frac{\partial z}{\partial r}\right) +\frac{\alpha ^{2}}{4}\frac{1}{%
\alpha }\frac{\partial }{\partial s}\left( \frac{1}{\alpha }\frac{\partial z%
}{\partial s}\right) &=&0  \nonumber \\
\frac{1}{4r^{3}}\frac{\partial }{\partial r}\left( r^{3}\frac{\partial z}{%
\partial r}\right) +\frac{1}{4}\frac{\partial ^{2}z}{\partial s^{2}} &=&0 
\nonumber
\end{eqnarray}

We can return to our problem. The equations%
\begin{eqnarray}
\frac{\partial I}{\partial z}+\frac{\partial }{\partial x}\left( vI\right) 
&=&0  \label{a49} \\
\frac{\partial v}{\partial z}+v\frac{\partial v}{\partial x} &=&c_{0}^{2}%
\frac{\partial }{\partial x}P\left[ I/I_{0}\right]   \nonumber
\end{eqnarray}%
where until now%
\begin{equation}
P\left[ I/I_{0}\right] =\frac{I}{I_{0}}  \label{a51}
\end{equation}%
and from now-on%
\begin{equation}
P\left[ I/I_{0}\right] =\frac{I}{I_{0}}-\beta \left( \frac{I}{I_{0}}\right)
^{K}  \label{a52}
\end{equation}%
and%
\begin{equation}
\frac{\partial }{\partial x}P\left[ I/I_{0}\right] =\frac{\delta P\left[
I/I_{0}\right] }{\delta \left( I/I_{0}\right) }\frac{\partial }{\partial x}%
\frac{I}{I_{0}}  \label{a53}
\end{equation}%
We intoduce the notation%
\begin{equation}
\frac{\delta P\left[ I/I_{0}\right] }{\delta \left( I/I_{0}\right) }\equiv G%
\left[ I\right]   \label{a56}
\end{equation}

The operations are the same as above. The first equation leads to%
\begin{equation}
\frac{\partial x}{\partial v}=v\frac{\partial z}{\partial v}-I\frac{\partial
z}{\partial I}  \label{a60}
\end{equation}

The second equation leads to%
\begin{equation}
\frac{\partial x}{\partial I}=G\left[ I\right] \frac{c_{0}^{2}}{I_{0}}\frac{%
\partial z}{\partial v}+v\frac{\partial z}{\partial I}  \label{a61}
\end{equation}%
and impose the equality of the mixed derivatives%
\begin{equation}
\frac{\partial }{\partial I}\left( v\frac{\partial z}{\partial v}-I\frac{%
\partial z}{\partial I}\right) =\frac{\partial }{\partial v}\left( G\left[ I%
\right] \frac{c_{0}^{2}}{I_{0}}\frac{\partial z}{\partial v}+v\frac{\partial
z}{\partial I}\right)  \label{a62}
\end{equation}%
\begin{equation}
v\frac{\partial ^{2}z}{\partial I\partial v}-\frac{\partial z}{\partial I}-I%
\frac{\partial ^{2}z}{\partial I^{2}}=G\left[ I\right] \frac{c_{0}^{2}}{I_{0}%
}\frac{\partial ^{2}z}{\partial v^{2}}+\frac{\partial z}{\partial I}+v\frac{%
\partial ^{2}z}{\partial I\partial v}  \label{a63}
\end{equation}%
We reduce the terms and obtain%
\begin{equation}
I\frac{\partial ^{2}z}{\partial I^{2}}+2\frac{\partial z}{\partial I}+G\left[
I\right] \frac{c_{0}^{2}}{I_{0}}\frac{\partial ^{2}z}{\partial v^{2}}=0
\label{a64}
\end{equation}%
As before the terms with derivatives to $I$ are grouped to give%
\begin{equation}
\frac{1}{I}\frac{\partial }{\partial I}\left( I^{2}\frac{\partial z}{%
\partial I}\right) +G\left[ I\right] \frac{c_{0}^{2}}{I_{0}}\frac{\partial
^{2}z}{\partial v^{2}}=0  \label{a65}
\end{equation}

In the first attempt we proceed in an analogous manner as above.

We make the substitution%
\begin{eqnarray}
r &=&I^{2}  \label{a69} \\
s &=&\frac{1}{2}\frac{v}{\sqrt{c_{0}^{2}/I_{0}}}  \nonumber
\end{eqnarray}%
We calculate%
\begin{eqnarray}
I &=&r^{2}  \label{a70} \\
\frac{\partial }{\partial r} &=&\frac{\partial }{\partial I}\frac{\partial I%
}{\partial r}=2r\frac{\partial }{\partial I}  \nonumber \\
\frac{\partial }{\partial I} &=&\frac{1}{2r}\frac{\partial }{\partial r} 
\nonumber
\end{eqnarray}%
\begin{eqnarray}
s &=&v\frac{1}{2}\frac{1}{\sqrt{c_{0}^{2}/I_{0}}}  \label{a71} \\
v &=&\alpha s\ \ \ \textrm{where}\ \ \alpha \equiv 2\sqrt{c_{0}^{2}/I_{0}} 
\nonumber \\
\frac{\partial }{\partial s} &=&\frac{\partial v}{\partial s}\frac{\partial 
}{\partial v}=\alpha \frac{\partial }{\partial v}  \nonumber \\
\frac{\partial }{\partial v} &=&\frac{1}{\alpha }\frac{\partial }{\partial s}
\nonumber
\end{eqnarray}%
This is replaced in the equation%
\begin{eqnarray}
\frac{1}{I}\frac{\partial }{\partial I}\left( I^{2}\frac{\partial z}{%
\partial I}\right) +G\left[ I\right] \frac{c_{0}^{2}}{I_{0}}\frac{\partial
^{2}z}{\partial v^{2}} &=&0  \label{a73} \\
\frac{1}{r^{2}}\frac{1}{2r}\frac{\partial }{\partial r}\left( r^{4}\frac{1}{%
2r}\frac{\partial z}{\partial r}\right) +G\left[ I\right] \frac{\alpha ^{2}}{%
4}\frac{1}{\alpha }\frac{\partial }{\partial s}\left( \frac{1}{\alpha }\frac{%
\partial z}{\partial s}\right) &=&0  \nonumber \\
\frac{1}{r^{3}}\frac{\partial }{\partial r}\left( r^{3}\frac{\partial z}{%
\partial r}\right) +G\left[ I\right] \frac{\partial ^{2}z}{\partial s^{2}}
&=&0  \nonumber
\end{eqnarray}%
The final form is%
\begin{equation}
\frac{\partial ^{2}z}{\partial r^{2}}+\frac{3}{r}\frac{\partial z}{\partial r%
}+G\left[ I\right] \frac{\partial ^{2}z}{\partial s^{2}}=0  \label{a78}
\end{equation}%
Here we must redefine $G$ as%
\begin{equation}
G\left[ I\right] \rightarrow G\left[ \sqrt{r}\right]  \label{a79}
\end{equation}

\bigskip

We make the substitution that combines the coordinate $r$ with the unknown
function $z$. [The coordinate $r$ is a measure of the intensity $I$.]%
\begin{equation}
\psi \equiv rz  \label{a81}
\end{equation}%
and replace the variable $t$ by $\psi $%
\begin{eqnarray}
z &=&\frac{\psi }{r}  \label{a83} \\
\frac{\partial z}{\partial r} &=&-\frac{1}{r^{2}}\psi +\frac{1}{r}\frac{%
\partial \psi }{\partial r}  \nonumber \\
\frac{\partial ^{2}z}{\partial r^{2}} &=&\frac{2}{r^{3}}\psi -\frac{1}{r^{2}}%
\frac{\partial \psi }{\partial r}-\frac{1}{r^{2}}\frac{\partial \psi }{%
\partial r}+\frac{1}{r}\frac{\partial ^{2}\psi }{\partial r^{2}}  \nonumber
\end{eqnarray}%
and we have%
\begin{eqnarray}
\frac{\partial ^{2}z}{\partial r^{2}} &=&\frac{\partial ^{2}}{\partial r^{2}}%
\left( \frac{\psi }{r}\right)   \label{a84} \\
&=&\frac{1}{r}\frac{\partial ^{2}\psi }{\partial r^{2}}-\frac{2}{r^{2}}\frac{%
\partial \psi }{\partial r}+\frac{2}{r^{3}}\psi   \nonumber
\end{eqnarray}%
\begin{eqnarray}
\frac{3}{r}\frac{\partial z}{\partial r} &=&\frac{3}{r}\frac{\partial }{%
\partial r}\left( \frac{\psi }{r}\right)   \label{a85} \\
&=&\frac{3}{r}\left( -\frac{1}{r^{2}}\psi +\frac{1}{r}\frac{\partial \psi }{%
\partial r}\right)   \nonumber
\end{eqnarray}%
and%
\begin{eqnarray}
G\left[ I\right] \frac{\partial ^{2}z}{\partial s^{2}} &=&G\left[ I\right] 
\frac{\partial ^{2}}{\partial s^{2}}\left( \frac{\psi }{r}\right) \ \ \textrm{%
remember }r\textrm{ and }s\textrm{ are independent}  \label{a86} \\
&=&G\left[ I\right] \frac{1}{r}\frac{\partial ^{2}\psi }{\partial s^{2}} 
\nonumber
\end{eqnarray}%
\begin{eqnarray}
&&\frac{1}{r}\frac{\partial ^{2}\psi }{\partial r^{2}}-\frac{2}{r^{2}}\frac{%
\partial \psi }{\partial r}+\frac{2}{r^{3}}\psi   \label{a90} \\
&&+\frac{3}{r}\left( -\frac{1}{r^{2}}\psi +\frac{1}{r}\frac{\partial \psi }{%
\partial r}\right)   \nonumber \\
&&+G\left[ r\right] \frac{1}{r}\frac{\partial ^{2}\psi }{\partial s^{2}} 
\nonumber \\
&=&0  \nonumber
\end{eqnarray}%
\begin{eqnarray}
\frac{1}{r}\frac{\partial ^{2}\psi }{\partial r^{2}}+\frac{1}{r^{2}}\frac{%
\partial \psi }{\partial r}-\frac{1}{r^{3}}\psi +G\left[ r\right] \frac{1}{r}%
\frac{\partial ^{2}\psi }{\partial s^{2}} &=&0  \label{a91} \\
\frac{\partial ^{2}\psi }{\partial r^{2}}+\frac{1}{r}\frac{\partial \psi }{%
\partial r}-\frac{1}{r^{2}}\psi +G\left[ r\right] \frac{\partial ^{2}\psi }{%
\partial s^{2}} &=&0  \nonumber
\end{eqnarray}%
For comparison that will allow identification of the operator we mention%
\begin{equation}
\Delta f\left( r,\varphi ,s\right) =\frac{1}{r}\frac{\partial }{\partial r}%
\left( r\frac{\partial f}{\partial r}\right) +\frac{1}{r^{2}}\frac{\partial
^{2}f}{\partial \varphi ^{2}}+\frac{\partial ^{2}f}{\partial s^{2}}
\label{a94}
\end{equation}%
We recognize the first two terms, containing the derivations to $r$. Then
the term $-\frac{1}{r^{2}}\psi $ can be attributed to the operator of
derivation with respect to the azimuthal variable $\varphi $%
\begin{equation}
\frac{1}{r^{2}}\frac{\partial ^{2}f}{\partial \varphi ^{2}}\rightarrow -%
\frac{1}{r^{2}}\psi   \label{a96}
\end{equation}%
if \cite{trubnikovbook}%
\begin{equation}
f\sim \psi \left( r,s\right) \cos \varphi   \label{a97}
\end{equation}%
Then our equation is%
\begin{equation}
\Delta _{\left( r,\varphi \right) }\Psi +G\left[ r\right] \frac{\partial
^{2}\Psi }{\partial s^{2}}=0  \label{a99}
\end{equation}%
where%
\begin{equation}
\Psi =\psi \cos \varphi   \label{a100}
\end{equation}%
and%
\begin{equation}
\psi =rz  \label{a101}
\end{equation}

\bigskip

Now we comment on the result of this derivation. We remember that the
variable $s$ comes from $v=\frac{\partial S}{\partial x}$ which is the
derivative of the eikonal to the transversal coordinate $x$.

If we introduce a harmonic variation on the $s$ coordinate%
\begin{equation}
\Psi =\Xi \left( r,\varphi \right) \exp \left( i\kappa _{s}s\right)
\label{a104}
\end{equation}%
we get a Helmholtz equation%
\begin{equation}
\Delta _{\left( r,\varphi \right) }\Xi \left( r,\varphi \right) \exp \left(
ik_{s}s\right) +G\left[ r\right] \left( -\kappa _{s}^{2}\right) \Xi \left(
r,\varphi \right) \exp \left( ik_{s}s\right) =0  \label{a105}
\end{equation}%
\begin{equation}
\left( \Delta _{\left( r,\varphi \right) }-\kappa _{s}^{2}G\left[ r\right]
\right) \Xi \left( r,\varphi \right) =0  \label{a106}
\end{equation}%
where 
\begin{equation}
\Xi \left( r,\varphi \right) =\psi \cos \varphi  \label{a107}
\end{equation}%
Here%
\begin{eqnarray}
\psi &=&rz  \label{a108} \\
&=&\left( \textrm{measure of the beam intensity, }I^{2}\right)  \nonumber \\
&&\times \left( \textrm{distance }z\textrm{ on axis}\right)  \nonumber
\end{eqnarray}%
and%
\begin{equation}
\varphi =\textrm{fictitious azimuthal angle}  \label{a109}
\end{equation}%
in a cylindrical space where the radial coordinate is $r=I^{2}$, the
vertical coordinate is $s\sim v\sim \frac{\partial S}{\partial x}$.

We note that instead of 
\begin{equation}
\kappa _{s}^{2}  \label{a110}
\end{equation}%
we now have%
\begin{equation}
G\left[ r\right] \kappa _{s}^{2}  \label{a111}
\end{equation}%
where $G$ decreases when the intensity increases. This means that the \emph{%
effective} wavenumber on the \textquotedblleft vertical\textquotedblright coordinate $s$ becomes smaller when $%
I$ increases. The rate of variation of $\Psi $ (which means $\psi =rz$)
along the direction $s$ becomes slower, with longer wavelengths along $s\sim 
\frac{\partial S}{\partial x}$.

In the absence of $G$ (\emph{i.e.} in the usual situation of self-focusing)
the two quantities $\psi =rz$ and $\partial S/\partial x$ evolve in a
similar way: $z$ increases approaching the focalization point, $z\rightarrow
z_{\ast }$. Simultaneously the intensity $I\sim \sqrt{r}$ increases hence $%
\psi \sim I^{2}z$ increases. The same is true for the derivative of the
eikonal since the field becomes more sharply concentrated on the transversal
coordinate $x$. Hence $\partial S/\partial x$ also increases when the beam
approaches focalization.

The explicit functional form of this correlated variation of the two
quantities $\psi $ and $\partial S/\partial x$ is difficult to be derived.
However we can see that by inserting $G\left[ r\right] $, which decreases
when the beam approaches focalization, it is affected the relative rate of
variation: $\psi $ will be slowed down along $s\sim \partial S/\partial x$
since $\left( \kappa _{s}^{2}\right) ^{eff}=G\left[ r\right] \kappa _{s}^{2}$
decreases as $G$. This is equivalent to slowing down the process of increase
of $I$ as approaching the focalization. The concentration of the energy of
the beam is slowed down. This is the manifestation of the well-known
physical process: increase of the density of electrons weakens the focusing
effect of the Kerr nonlinearity and the focalization saturates.

\bigskip

\section{Appendix B. Estimation of the physical parameters}

\label{App:AppendixB}

\renewcommand{\theequation}{B.\arabic{equation}} \setcounter{equation}{0}

\subsection{Estimation of the diffusion coefficient}

The distance travelled by an electron between two collisions is 
\begin{equation}
\delta \sim v_{th,e}\times \tau  \label{b1}
\end{equation}%
where the thermal velocity must correspond to few electron-volts since the
electrons are just after being created with $E_{g}=11\ \left( eV\right) $
and then heated. We take%
\begin{equation}
E^{elect}\sim 1\ eV  \label{b2}
\end{equation}%
and the thermal velocity%
\begin{eqnarray}
v_{th,e} &=&4.19\times 10^{5}\ \sqrt{T_{e}\left( eV\right) }\ \left( \frac{m%
}{s}\right)  \label{b3}
\end{eqnarray}%
and the time of collisions%
\begin{equation}
\tau =10^{-13}\ \left( s\right)  \label{b4}
\end{equation}%
Then the distance between two collisions%
\begin{eqnarray}
\delta &\sim &v_{th,e}\times \tau  \label{b6} \\
&=&4\times 10^{5}\times 10^{-13}  \nonumber
\end{eqnarray}
is of the order of $10^{-7}$.

On the other hand we have an alternative estimation%
\begin{equation}
\tau =\textrm{collision time}=\nu _{ee}^{-1}  \label{b7}
\end{equation}%
and%
\begin{equation}
\nu _{ee}=2.91\times 10^{-6}\ \ln \Lambda \times \frac{n_{e}\left(
cm^{-3}\right) }{\left[ T_{e}\left( eV\right) \right] ^{3/2}}\ \ \ \ \ \ \ \
\ \left( s\right)  \label{b8}
\end{equation}%
Take%
\begin{eqnarray}
n_{e} &\sim &10^{23}\ \left( m^{-3}\right) =10^{17}\ \left( cm^{-3}\right)
\label{b9} \\
T_{e} &\sim &1\ \left( eV\right)  \nonumber \\
\ln \Lambda &=&25  \nonumber
\end{eqnarray}%
It results%
\begin{eqnarray}
\nu _{ee} &=&3\times 10^{-6}\times 25\times \frac{10^{17}}{\left[ 1\right]
^{3/2}}  \label{b10} \\
&=&75\times 10^{11}\ \left( s^{-1}\right)  \nonumber
\end{eqnarray}%
An order of magnitude is
\begin{equation}
\nu _{ee}\sim 10^{13}\ \ \left( s^{-1}\right)  \label{b11}
\end{equation}%
this is compatible with 
\begin{eqnarray}
\tau &\sim &\nu _{ee}^{-1}  \label{b16} \\
\tau &=&10^{-13}\ \left( s\right)  \nonumber
\end{eqnarray}
which is compatible with Ref.\cite{mlejnek98}.

\bigskip

If we use as input the frequency of collisions $\tau $ and calculate the
temperature of the electron plasma%
\begin{equation}
\nu _{ee}=\tau ^{-1}=10^{13}\ \left( s^{-1}\right)  \label{b17}
\end{equation}%
\begin{eqnarray}
T_{e}^{3/2} &=&\frac{2.91\times 10^{-6}\times \ln \Lambda \times n_{e}}{\nu
_{ee}}  \label{b19} \\
&=&\frac{3\times 10^{-6}\times 25\times 10^{17}}{10^{13}}=75\times 10^{-2} 
\nonumber
\end{eqnarray}%
it results%
\begin{eqnarray}
T_{e} &=&\left( 75\times 10^{-2}\right) ^{2/3}  \label{b20} \\
&\approx&0.8\ \ \left( eV\right)  \nonumber
\end{eqnarray}

compatible with our assumtion.

We can estimate the energy that can go to the plasma of electrons.%
\begin{equation}
P_{in}\sim 10^{9}\ \left( W\right)  \label{b21}
\end{equation}%
For this we introduce a parameter $fraction$ that represents the amount from
the total energy that goes to the electron plasma. The energy is%
\begin{eqnarray}
W^{elect-plasma} &=&fraction\times P_{in}\times \Delta t  \label{b22} \\
&=&1\times 10^{9}\times 100\ \left( fs\right)  \nonumber \\
&=&10^{9}\times 10^{-13}  \nonumber \\
&=&10^{-4}\ \left( J\right)  \nonumber
\end{eqnarray}%
This energy is distributed on a number of particles $N$%
\begin{eqnarray}
N &=&\rho \times Vol  \label{b23} \\
&=&10^{23}\ \left( m^{-3}\right) \times a^{3}  \nonumber
\end{eqnarray}%
where 
\begin{equation}
a\sim 1\ \left( mm\right) =10^{-3}  \label{b24}
\end{equation}%
\begin{equation}
N=10^{23}\times 10^{-9}=10^{14}\ \ \left( particles\right)  \label{b245}
\end{equation}%
The amount of energy for each particle (electron) is%
\begin{eqnarray}
\delta W^{elect-plasma} &=&\frac{W^{elect-plasma}}{N}  \label{b30} \\
&=&\frac{10^{-4}\ \left( J\right) }{10^{14}\ \left( electrons\right) } 
\nonumber \\
&=&10^{-18}\ \left( J\right)  \nonumber
\end{eqnarray}%
This energy correspnds to%
\begin{eqnarray}
T_{e}^{elec} &\sim &\frac{\delta W^{elect-plasma}}{\left( eV\right) }=\frac{%
10^{-18}\ \left( J\right) }{1.6\times 10^{-19}\ \left( J/eV\right) }
\label{b32} \\
&\sim &10\ \left( eV\right)  \nonumber
\end{eqnarray}%
we have%
\begin{eqnarray}
\frac{m_{e}v_{th,e}^{2}}{2} &=&\delta W^{elect-plasma}  \label{b35} \\
v_{th,e}^{2} &=&\frac{2\delta W^{elect-plasma}}{m_{e}}=\frac{2\times
10^{-18}\ \left( J\right) }{9.1\times 10^{-31}\ \left( kg\right) }  \nonumber
\\
&=&0.2\times 10^{13}\ \left( \frac{J}{kg}\right)  \nonumber
\end{eqnarray}%
\begin{equation}
v_{th,e}=1.4\times 10^{6}\ \left( \frac{m}{s}\right)  \label{b36}
\end{equation}%
The distance traversed in a time $\tau =10^{-13}\ \left( s\right) $ is%
\begin{equation}
\delta =v_{th,e}\times \tau =10^{6}\times 10^{-13}=10^{-7}\ \left( m\right)
\label{b37}
\end{equation}

Exactly the same result as above.

\bigskip

\begin{eqnarray}
\delta &\sim &10^{-6}\ \ \left( m\right) \ \ \textrm{rather arbitrary}
\label{b38} \\
\tau &\sim &1\times 10^{-13}\ \left( s\right) \textrm{\ \ according to Mlejnek}
\nonumber \\
d &\sim &\frac{10^{-12}}{10^{-13}}=10\ \ \left( \frac{m^{2}}{s}\right) 
\nonumber
\end{eqnarray}%
Possibly the range of the diffusion coefficient would be%
\begin{equation}
d\in \left[ 0.1,10\right] \ \left( \frac{m^{2}}{s}\right)  \label{b39}
\end{equation}

We choose%
\begin{equation}
d=10\ \left( \frac{m^{2}}{s}\right)  \label{b40}
\end{equation}

\subsection{Estimation of the effect of focusing and defocusing terms}

We will use%
\begin{eqnarray}
n_{2} &=&3.2\times 10^{-19}\ \left( \frac{cm^{2}}{W}\right) \ \ \left( \textrm{Ref.\cite%
{skupinintense}}\right)  \label{b41} \\
&=&3.2\times 10^{-23}\ \left( \frac{m^{2}}{W}\right)  \nonumber
\end{eqnarray}%
\begin{equation}
\sigma \sim 5.1\times 10^{-24}\left( m^{2}\right) \ \ \ \left( \textrm{Ref.\cite{mlejnek98}}%
\right)  \label{b45}
\end{equation}%
\begin{equation}
\tau _{0}\sim {\small 3.5\times 10}^{-13}{\small \ }\left( s\right) \ \
\left( \textrm{Ref.\cite{mlejnek98}}\right)  \label{b46}
\end{equation}%
\begin{equation}
\rho =10^{23}\ \left( m^{-3}\right)  \label{b47}
\end{equation}%
\begin{equation}
I_{0}\sim 10^{15}\ \left( \frac{W}{m^{2}}\right) ...10^{17}\ \left( \frac{W}{%
m^{2}}\right)  \label{b48}
\end{equation}%
This is intensity on the whole area. In spots where self-focalization takes
place, it can be orders of magnitude higher.%
\begin{equation}
\lambda =775\ \left( nm\right)  \label{b62}
\end{equation}%
From the last data%
\begin{eqnarray}
k_{0} &=&\frac{2\pi }{\lambda }=\frac{2\pi }{775\times 10^{-9}\left(
m\right) }=\frac{2\pi }{0.775\times 10^{-6}\left( m\right) }  \label{b63} \\
&\sim &8\times 10^{6}\ \left( m^{-1}\right)  \nonumber
\end{eqnarray}%
\begin{eqnarray}
\frac{\omega _{0}}{k_{0}} &=&c  \label{b64} \\
\omega _{0} &=&k_{0}c=8\times 10^{6}\left( \frac{1}{m}\right) \times 3\times
10^{8}\ \left( \frac{m}{s}\right) =24\times 10^{14}\ \left( s^{-1}\right) 
\nonumber
\end{eqnarray}%
Then%
\begin{eqnarray}
\frac{2}{c}n_{2} &=&2\times \frac{1}{3\times 10^{8}\left( \frac{m}{s}\right) 
}\times 10^{-23}\left( \frac{m^{2}}{W}\right)  \label{b65} \\
&=&0.6\times 10^{-31}\ \left( \frac{ms^{2}}{J}\right)  \nonumber
\end{eqnarray}%
from where%
\begin{eqnarray}
\alpha &\equiv &\frac{2\omega _{0}}{c}n_{2}=\left( \omega _{0}\right) \times 
\frac{2}{c}n_{2}  \label{b66} \\
&\sim &24\times 10^{14}\left( \frac{1}{s}\right) \times 0.6\times
10^{-31}\left( \frac{ms^{2}}{J}\right)  \nonumber \\
&=&1.44\times 10^{-16}\ \left( \frac{m}{W}\right)  \nonumber
\end{eqnarray}

\bigskip

The constant in the defocusing term%
\begin{eqnarray}
\sigma \tau _{0} &\sim &5\times 10^{-24}\left( m^{2}\right) \times 3.5\times
10^{-13}\left( s\right)  \label{b67} \\
&=&1.75\times 10^{-36}\ \left( m^{2}s\right)  \nonumber
\end{eqnarray}%
from where%
\begin{eqnarray}
\alpha ^{\prime } &\equiv &\omega _{0}\sigma \tau _{0}  \label{b68} \\
&\sim &24\times 10^{14}\left( \frac{1}{s}\right) \times 1.75\times 10^{-36}\
\left( m^{2}s\right)  \nonumber \\
&=&4.2\times 10^{-21}\ \left( m^{2}\right)  \nonumber
\end{eqnarray}

Now we can estimate the two terms that compete%
\begin{equation}
\alpha I^{2}-\alpha ^{\prime }\rho I  \label{b72}
\end{equation}%
factorizing a $I$ we have%
\begin{eqnarray}
&&\alpha I-\alpha ^{\prime }\rho  \label{b73} \\
&\sim &1.44\times 10^{-16}\ \left( \frac{m}{W}\right) \times 10^{15}\ \left( 
\frac{W}{m^{2}}\right) -4.2\times 10^{-21}\ \left( m^{2}\right) \times
10^{23}\ \left( \frac{1}{m^{3}}\right)  \nonumber \\
&=&0.144\ \left( \frac{1}{m}\right) -420\left( \frac{1}{m}\right)  \nonumber
\end{eqnarray}

If instead of $I\sim 10^{15}\ \left( \frac{W}{m^{2}}\right) $ we would have
taken 
\begin{equation}
I\sim 10^{17}\ \left( \frac{W}{m^{2}}\right)  \label{b74}
\end{equation}%
Then%
\begin{eqnarray}
&&\alpha I-\alpha ^{\prime }\rho  \label{b75} \\
&\sim &14-420  \nonumber
\end{eqnarray}%
and the two terms were closer, with still huge dominance of the second term,
which represents \emph{defocusing} due to plasma, over the \emph{focusing}
term due to Kerr nonlinearity.

However in the spots of focalization, which develop spontaneously in a strip
of high $I$, the local intensity is higher. Then the focalization overcomes
the defocusing action of the electrons.

\emph{It looks that we must work at the limit of balance of the focusing and
defocusing, with a certain dominance of the Kerr-induced focusing, since we
want to study the displacement of the front and motion of the interface
associated with the relocation of the high-}$I$ \emph{zone.}

\bigskip

We conclude after using the usual values of the parameters \cite%
{skupinintense}, \cite{mlejnek98}  
\begin{eqnarray}
\alpha  &\sim &1.44\times 10^{-16}\ \left( \frac{m}{W}\right)   \label{b99}
\\
\alpha ^{\prime } &\sim &4.2\times 10^{-21}\ \left( m^{2}\right)   \nonumber
\end{eqnarray}%
and may be used with%
\begin{eqnarray}
I &\sim &10^{17}\ \left( \frac{W}{m^{2}}\right) \ \ \left( \textrm{or higher}%
\right)   \label{b110} \\
\rho  &\sim &10^{23}\ \left( \frac{1}{m^{3}}\right)   \nonumber
\end{eqnarray}

\bigskip

\bigskip

For recombination%
\begin{equation}
a=5\times 10^{-13}\ \left( \frac{m^{3}}{s}\right)   \label{b120}
\end{equation}%
and for MPI $\beta ^{\left( K=7\right) }=6.5\times 10^{-104}\ \left( \frac{%
m^{11}}{W^{6}}\right) $ we have%
\begin{equation}
b\equiv \frac{\beta ^{\left( K=7\right) }}{K\hslash \omega _{0}}=3.6\times
10^{-86}\ \ \ \left( \frac{m^{11}}{J}\right)   \label{b130}
\end{equation}%
Taking $E^{phys}=9.15\times 10^{7}\ \left( \frac{V}{m}\right) $ we obtain $%
I\equiv \left\vert \widetilde{E}_{0}\right\vert ^{2}\ \ \left( \frac{W}{m^{2}%
}\right) $, alternatively $I=\left( \left\vert \sqrt{c\varepsilon _{0}}%
E^{phys}\right\vert ^{2}\right) $ such that, calculated below, we have for $%
K=7$%
\begin{equation}
bI^{K}\sim 7.7\times 10^{7}\ \left( \frac{1}{m^{3}s}\right)   \label{b140}
\end{equation}

These are the values of the parameters that are used in the main text.

\end{appendices}


\end{document}